\documentclass{ws-procs10x7}

\usepackage{amssymb}  

\let\cropmarks\relax

\def\preprint{%
\hbox to \textwidth{%
{\large\sffamily\bfseries University of Wisconsin - Madison}\hfill
\vtop{%
\normalsize
\hbox{\sffamily\bfseries MADPH-05-1444}
\hbox{\bf hep-ph/0510276}
\hbox{\hfil}
}}
}

\begin{document}

\title{Lepton Photon Symposium 2005: Summary and Outlook$^*$}
\author{Francis Halzen}
\address{Physics Department, University of Wisconsin, Madison, WI 53706, USA\\
Email: \ halzen@pheno.physics.wisc.edu}

\twocolumn[\maketitle\abstract{Lepton Photon 2005 told the saga of the Standard Model which is still exhilarating because it leaves all questions of consequence unanswered.}]

\vspace*{-2.35in}

\preprint

\vspace{1.7in}

\footnotetext{%
$^*$Presented at the {\it XXII International Symposium on Lepton-Photon Interactions at High Energy}, Uppsala, Sweden, July 2005.}

\thispagestyle{empty}

 Over the last decade the biennial gathering discussing leptons and photons has broadened its horizons to reflect the excursions particle physics techniques have made into astronomy and cosmology. It was in the grandest of particle physics traditions however  that five days of talks in the historic aula of one of Europe's oldest universities, the home of Linnaeus, Manne and Kai Siegbahn and Dag Hammarskjold, erected the impressive edifice that is called the Standard Model. Experimental ingenuity has not been able to pierce the Model's armor and I cannot help thinking of the prophetic words of Leon Lederman at the Rochester meeting held in Madison twenty five years ago: ``the experimentalists do not have enough money and the theorists are overconfident". Where  experimentalists are concerned, nobody could have anticipated that today we would be studying the proton structure to one thousandth its size and would have established the Standard Model as a gauge theory with a precision of one in a thousand, pushing any interference of possible new physics to energy scales beyond 10 TeV. The theorists can modestly claim that they have taken revenge for Leon's remark. Because all the big questions remain unanswered, there is no feeling though that we are now dotting the i's and crossing the t's of a mature theory. Worse, the theory has its own demise built into its radiative corrections.
 
The most evident of unanswered questions is why the weak interactions are weak. Though unified with electromagnetism, electromagnetism is apparent in daily life while the weak interactions are not. Already in 1934 Fermi provided an answer with a theory\cite{Fermi} that prescribed a quantitative relation between the fine-structure constant and the weak coupling $G \sim \alpha /  m_W^2$. Although Fermi adjusted $m_W$ to accommodate the strength and range of nuclear radioactive decays, one can readily obtain a value of $m_W$ of 40 GeV from the observed decay rate of the muon for which the proportionality factor is $\pi \over \sqrt2$. The answer is off by a factor of 2 because the discovery of parity violation and neutral currents was in the future and introduces an additional factor $1-  m_W^2 / m_Z^2$:
\begin{equation}
G_\mu =\left[\pi \alpha \over \sqrt2 m_W^2\right]\, \left[1 \over {1 - m_W^2 / m_Z^2}\right]\, ( 1 + \Delta r)\,.
\label{Fermi}
\end{equation}
Fermi could certainly not have anticipated that we now have a renormalizable gauge theory that allows us to calculate the radiative corrections $\Delta r$ to his formula. Besides regular higher order diagrams, loops associated with the top quark and the Higgs boson contribute; they have been observed\cite{sd,aj,dd}.

I once heard one of my favorite physicists refer to the Higgs as the ``ugly" particle, but this is nowadays politically incorrect. Indeed, scalar particles are unnatural.  If one calculates the radiative corrections to the mass $m$ appearing in the Higgs potential, the same gauge theory that withstood the onslaught of precision experiments at LEP/SLC and the Tevatron yields a result that grows quadratically:
\begin{equation}
\delta m^2 = {3\over 16\pi^2 v^2} (2m_W^2 + m_Z^2 + m_H^2 - 4 m_t^2) \Lambda^2, 
\label{quadrdiv}
\end{equation}
where $m_H^2=2\lambda v^2$, $\lambda$ is the quartic Higgs coupling, $v=246$ GeV and $\Lambda$ a cutoff. Upon minimization of the potential, this translates into a dangerous contribution to the Higgs vacuum expectation value which destabilizes the electroweak scale\cite{casas}. The Standard Model works amazingly well by fixing $\Lambda$ at the electroweak scale. It is generally assumed that this indicates the existence of new physics beyond the Standard Model; following Weinberg
\begin{eqnarray}
{\cal L}(m_W) &=& {1\over 2} m^2 H^\dagger H + {1 \over 4} \lambda(H^\dagger H)^2 + {\cal L}^{\rm gauge}_{\rm SM} 
 \nonumber\\
&+&  {\cal L}^{\rm Yukawa}_{\rm SM} + {1\over\Lambda}{\cal L}^5 +  {1\over\Lambda^2}{\cal  L}^6+ ... .
\label{weinberg}
\end{eqnarray} 
The operators of higher dimension param\-etrize physics beyond the Standard Model. The optimistic interpretation of all this is that, just like Fermi anticipated particle physics at 100\,GeV in 1934, the electroweak gauge theory requires new physics to tame the divergences associated with the Higgs potential.  By the most conservative estimates this new physics is within our reach. Avoiding fine-tuning requires $\Lambda \lesssim 2{\sim}3$\,TeV to be revealed by the LHC, possibly by the Tevatron. For instance, for $m_H = 115$--200~GeV
\begin{equation}
\label{quadrft}
\left|{\delta m^2 \over m^2}\right|=
\left|{\delta v^2 \over v^2}\right|\leq 10\ \Rightarrow \ 
\Lambda  \lesssim 2\mbox{--}3\ {\rm TeV} \, .
\end{equation}

Dark clouds have built up around this sunny horizon because some electroweak precision measurements match the Standard Model predictions with too high precision, pushing $\Lambda$ to 10\,TeV. The data pushes some of the higher order dimensional operators in Weinberg's effective Lagrangian to scales beyond 10\,TeV.  Some theorists have panicked by proposing that the factor multiplying the unruly quadratic correction $(2  m_W^2+m_Z^2+ m_H^2-4m_t^2)$ must vanish; exactly! This has been dubbed the Veltman condition. The problem is now ``solved" because scales as large as 10 TeV, possibly even higher, can be accommodated by the observations once one eliminates the dominant contribution. One can even make this stick to all orders and for  $\Lambda \leq 10$\,TeV, this requires that $m_H \sim 210$--225\,GeV\cite{casas}.

Let's contemplate the possibilities. The Veltman condition happens to be satisfied and this would leave particle physics with an ugly fine tuning problem reminiscent  of the cosmological constant. This is very unlikely; LHC must reveal the ÒHiggsÓ physics already observed via radiative correction, or at least discover the physics that implements the Veltman condition\cite{LHC}. It must appear at $2{\sim}3$ TeV, even though higher scales can be rationalized when accommodating selected experiments\cite{sd}. Minimal supersymmetry is a textbook example. Even though it elegantly controls the quadratic divergence by the cancellation of boson and fermion contributions, it is already fine-tuned at a scale of $2 \sim 3$\,TeV. There has been an explosion of creativity to resolve the challange in other ways; the good news is that all involve new physics in the form of scalars, new gauge bosons, non-standard interactions\dots\ Alternatively, it is possible that we may be guessing the future while holding too small a deck of cards and LHC will open a new world that we did not anticipate. Particle physics would return to its early traditions where experiment leads theory, as it should be, and where innovative techniques introduce new accelerators and detection methods that allow us to observe with an open mind and without a plan, leading us to unexpected discoveries.

\begin{figure*}[t]
\centering\leavevmode
\epsfig{file=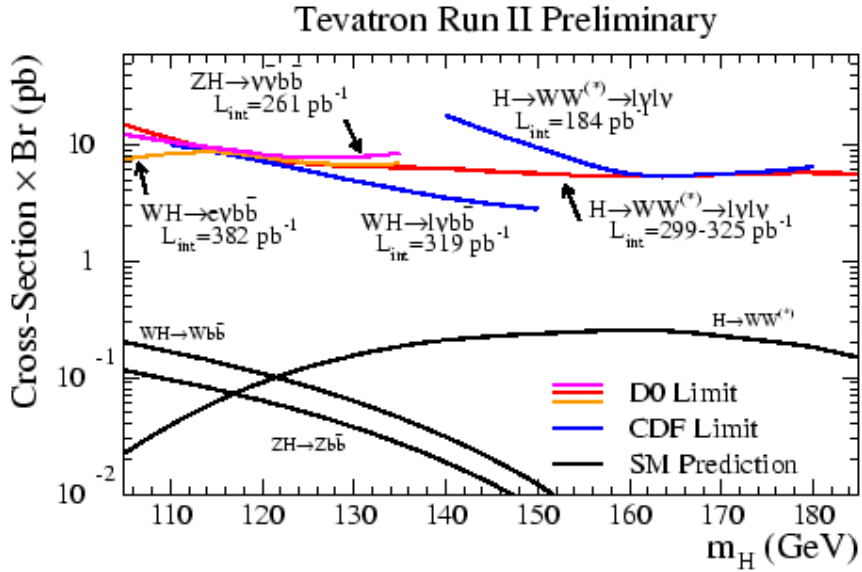, width=4.5in}
\caption{The Tevatron roadmap to the Higgs by increased luminosity and improved detector performance.}
\label{fig:tevatron}
\end{figure*}

There is good news from Fermilab\cite{sl}. The Tevatron experiments are within an order of magnitude of the sensitivity where they may discover the Higgs; see Fig.\,\ref{fig:tevatron}. The integrated luminosity of 8\,fb$^{-1}$, expected by extrapolating present collider performance, bridges that gap. Discovery will require an additional boost in sensitivity from improved detector performance which is actually expected for lepton identification and jet mass resolution. The performance of the detectors has been nothing short of spectacular as illustrated by the identification of the top quark in 6-jet events\cite{aj}.

\begin{figure*}[t]
\centering\leavevmode
\epsfig{file=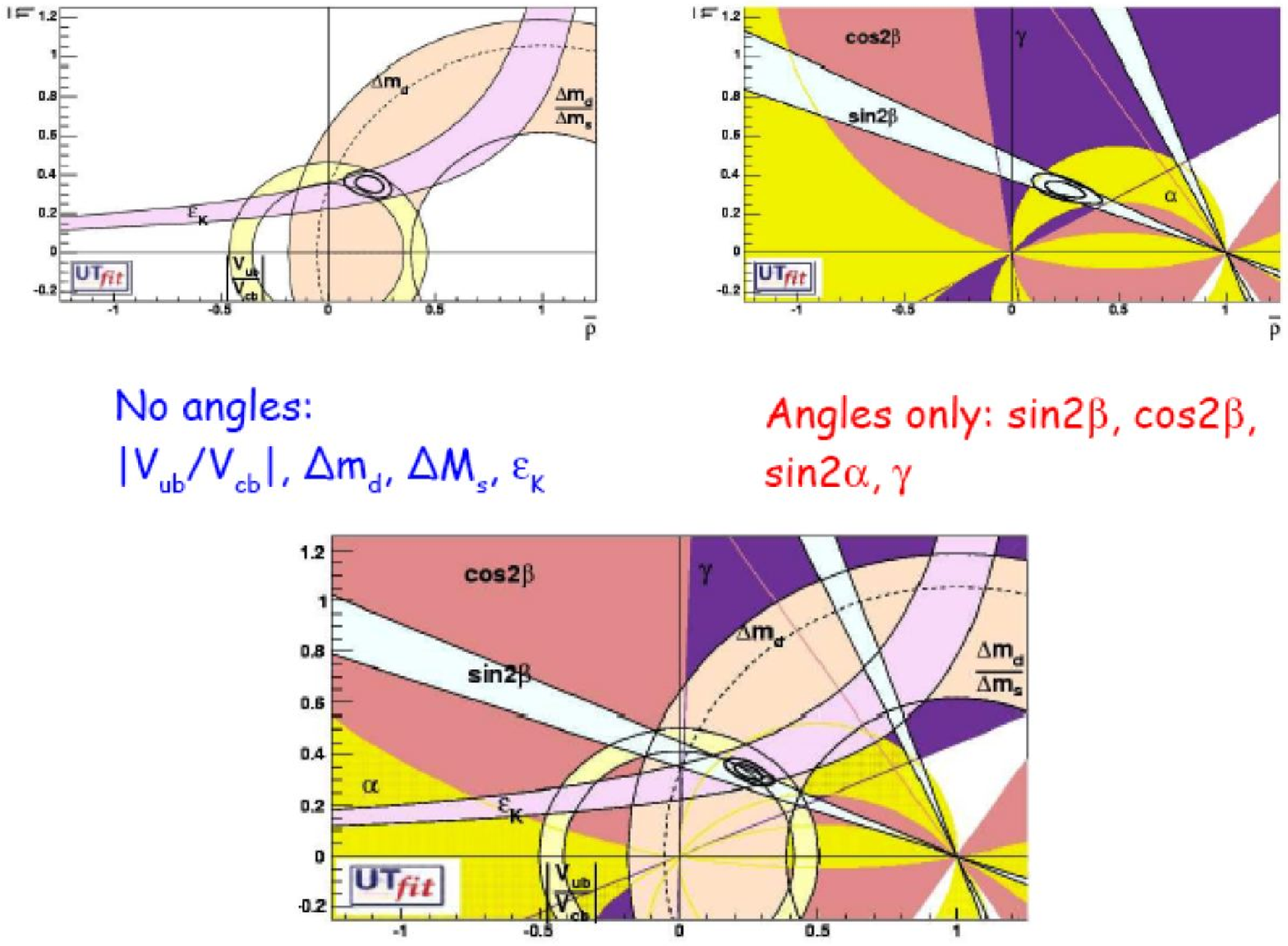, width=5.5in}
\caption{Geometry of the CKM triangle converges on the standard model by measuring sides, angles, or both.}
\label{fig:hiprecision}
\end{figure*}

Baryogenesis is another one of the grand issues left unresolved by the Standard Model. We know that at some early time in the evolution of the Universe quarks and antiquarks annihilated into light, except for just one quark in $10^{10}$ that failed to find a partner and became us. We are here because baryogenesis managed to accommodate the three Zakharov conditions; one of them dictates CP-violation. Evidence for the indirect violation of CP-invariance was first revealed in 1964 in the mixing of neutral kaons. Direct CP-violation, not mixing-assisted, was not discovered until 1999. Today, precision data on neutral kaons have been accumulated over 40 years; the measurements can, without exception, be accommodated by the Standard Model with three families\cite{eb}. History has repeated itself for $B$ mesons, but in three years only, thanks to the magnificent performance of the $B$-meson factories Belle and BaBar\cite{bb}. Direct CP-violation has been established in the decay $B_d\rightarrow K\pi$ with a significance in excess of 5 sigma. Unfortunately, this result, as well as a wealth of data contributed by CLEO, BES and Dafne, fails to reveal evidence for new physics\cite{c}. Whenever the experimental precision increases, the higher precision measurements invariably collapse onto the Standard Model values; see Fig.\,\ref{fig:hiprecision}. Given the rapid progress and the better theoretical understanding of the Standard Model expectations relative to the $K$ system\cite{ns}, the hope is that at this point the glass is half full and that improved data will pierce the Standard Model's resistant armor\cite{s}. Where theory is concerned, it is noteworthy that lattice techniques have reached the maturity to perform computer experiments that are confirmed by experiment.

The rise and fall of theories, or at least of their popularity, can be easily assessed by consulting the citation index. The number of citations to Wolfenstein's seminal paper on neutrino oscillations in the presence of matter has, after a steady increase from 1978--2000, dropped by almost a factor of two since. Progress in neutrino physics has been led by a string of fundamental experimental measurements summarized by the simple vacuum relations between the neutrino states produced by the weak interactions in e, mu and tau flavors and propagating as mixed states $\nu_1$, $\nu_2$ and~$\nu_3$:
\begin{eqnarray}
\nu_1 &=& -\cos\theta\, \nu_e + \sin\theta \left(\nu_\mu-\nu_\tau\over\sqrt2\right) \,, \nonumber \\
\nu_2 &=&  \sin\theta\, \nu_e +  \cos\theta\left(\nu_\mu-\nu_\tau\over\sqrt2\right) \,, \nonumber \\
\nu_3  &=& \left( \nu_\mu + \nu_\tau \over \sqrt2 \right)\,. \label{mixing}
\end{eqnarray}
Here $\theta$ is the solar mixing angle. Discovery of neutrino oscillations in solar and atmospheric beams has been confirmed by supporting evidence from reactor and accelerator beams\cite{nu}.

As usual, next-generation experiments are a lot more challenging and the boom times of neutrino physics are probably over as reflected by Wolfenstein's citations. Also, high-precision data from the pioneering experiments trickle in at a slower pace, although new evidence for the oscillatory behavior in $L/E$ of the muon-neutrinos in the atmospheric neutrino beam has become very convincing. The new results included first data from the reborn SuperKamiokande experiment\cite{ys}. The future of neutrino physics is undoubtedly bright. Construction of the KATRIN spectrometer measuring neutrino mass to 0.2 eV by studying the kinematics of tritium decay is in progress and a wealth of ideas on double beta decay and long-baseline experiments is approaching reality\cite{cw}. These experiments will have to answer the great ``known-unknowns" of neutrino physics: their absolute mass and hierarchy, the precise value of the second and third small mixing angle and its associated CP-violating phase and whether neutrinos are really Majorana particles. In Eq.~(\ref{mixing}) we assumed that the mixing of mu and tau neutrinos is maximal, with no admixture of electron neutrinos. Observing otherwise will most likely require next-generation experiments.

Among these, discovery of neutrinoless double beta decay would be especially rewarding\cite{oc}. Its observation would confirm the theoretical bias that neutrinos are their own antiparticles, yield critical information on the absolute mass scale and, possibly, resolve the hierarchy problem. In the meantime we will keep wondering whether small neutrino masses are our first glimpse at grand unified theories via the see-saw mechanism, or represent a new Yukawa scale tantalizingly connected to lepton conservation and, possibly, the cosmological constant. 

The cosmological constant represents a thorny issue for the Standard Model\cite{ewit}. New physics is also required to control the Standard Model calculation of the vacuum energy, also known as the cosmological constant, which diverges as
\begin{equation}
\int^\Lambda  {1\over2}\hbar\omega = \int^\Lambda {1\over2}\hbar\sqrt{k^2+ m^2\,}\,  d^2k \ \sim \ \Lambda^4 \,.
\end{equation}
It has not escaped attention that the cutoff energy required to accommodate its now ``observed'' value happens to be $\Lambda = 10^{-3}$\,eV, of the order of the neutrino mass.

\begin{figure}[h]
\centering\leavevmode
\epsfig{file=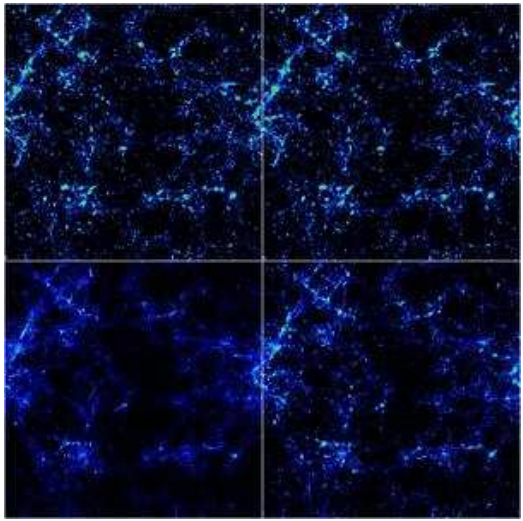, width=2.7in}
\caption{Simulations of structure formation with varying amounts of matter in the neutrino component, {\it i.e.}\ varying neutrino mass: (top left) $m_\nu = 0$~eV; (top right) $m_\nu = 1$~eV; (bottom right) $m_\nu = 4$~eV;  (bottom left) $m_\nu = 7$~eV.}
\label{fig:numass}
\end{figure}

Information on neutrino mass has\break
 emerged from an unexpected direction: cosmology\cite{sh}. The structure of the Universe is dictated by the physics of cold dark matter and the galaxies we see today are the remnants of relatively small overdensities in the nearly uniform distribution of matter in the very early Universe.
 Overdensity means overpressure that drives an acoustic wave into the other components making up the Universe: the hot gas of nuclei and photons and the neutrinos. These acoustic waves are seen today in the temperature fluctuations of the microwave background as well as in the distribution of galaxies on the sky.  With a contribution to the Universe's matter balance similar to that of light, neutrinos play a secondary role. The role is however identifiable --- neutrinos, because of their large mean-free paths, prevent the smaller structures in the cold dark matter from fully developing and this is visible in the observed distribution of galaxies; see Fig.\,\ref{fig:numass}. Simulations of structure formation with varying amounts of matter in the neutrino component, {\it i.e.}\ varying neutrino mass, can be matched to a variety of observations of today's sky, including measurements of galaxy-galaxy correlations and temperature fluctuations on the surface of last scattering. The results suggest a neutrino mass of at most 1 eV, summed over the 3 neutrino flavors, a range compatible with the one deduced from oscillations.
 
Cosmology, in association with the discovery of neutrino mass, has also been responsible for renewed interest in deciphering baryogenesis --- a tally of the rapidly increasing number of citations to the 1986 paper by Fukugita and Yanagida underscores the point. The problem is more clearly framed than ever before. The imprint on the surface of last scattering of the acoustic waves driven into the hot gas of nuclei and photons reveals a relative abundance of baryons to photons of $6.5^{ +0.4}_{-0.3} \times10^{-10}$ (WMAP observation). Gamov realized that a Universe born as hot plasma must consist mostly of hydrogen and helium, with small amounts of deuterium and lithium added. The detailed balance depends on basic nuclear physics as well as the same relative abundance of baryons to photons; the state of the art result of this exercise yields $4.7^{+1.0}_{-0.8}\times 10^{-10}$.  The agreement of the two observations is stunning, not just because of the precision, but because of the concordance of two results derived by totally unrelated ways to probe the early Universe.

Physics at the high energy frontier is the physics of partons. For instance, at the LHC gluons produce the Higgs boson and the highest energy neutrinos interact with sea-quarks in the detector. We master this physics with unforeseen precision because of a decade of steadily improving HERA measurements of the nucleon structure\cite{hera}. These now include experiments using targets of polarized protons and neutrons. HERA is our nucleon microscope, tunable by the wavelength and the fluctuation time of the virtual photon exchanged in the electron proton collision. The wavelength of the virtual photons probing the nucleon is reduced with increased momentum transfer $Q$. The proton has now been probed to distances of one thousandth of its size of 1 fm. In the interaction the fluctuations of the virtual photons survive over distances $ ct \sim1/ x$, where $x$ is the relative momentum of the parton. HERA now studies the production of chains of gluons as long as 10 fm, an order of magnitude larger than and probably totally insensitive to the proton target. These are novel QCD structures, the understanding of which has been challenging\cite{gi}. We should not forget however that theorists analyze HERA data with calculations performed to next-to-next to leading order in the strong coupling. In fact, beyond this precision one has to include the photon as a parton inside the proton\cite{nutev}. These electromagnetic structure functions violate isospin and differentiate a $u$-quark in a proton from a $d$-quark in a neutron because of the different electric charge of the quark. Interestingly, their inclusion in the structure functions modifies the extraction of the Weinberg angle from NuTeV data, bridging roughly half of its discrepancy with the particle data book value.  Added to already anticipated intrinsic isospin violations associated with sea-quarks, the NuTeV anomaly may be on its way out.

Recalling Lederman, whatever the actual funding, the experimenters managed to deliver most highlights of this conference. And where history has proven that theorists had the right to be confident in 1980, they have not faded into the background, and provided some highlights of their own. Developing QCD calculations to the level that the photon structure of the proton becomes a factor is a tour de force and, there were others at this meeting. Progress in higher order QCD computations of hard processes is mind boggling --- progress useful, sometimes essential, for the interpretation of LHC experiments\cite{gs}. Discussions of strings, supersymmetry and additional dimensions were very much focused on the capability of experiments to confirm or debunk these concepts\cite{ia}.

Theory and experiment joined forces in the ongoing attempts to read the information supplied by the data on heavy ion collisions from Brookhaven. Rather than the anticipated quark gluon plasma, the data suggests the formation of a strongly interacting fluid with very low viscosity for its entropy\cite{js}. Similar fluids of cold $^6$Li atoms have been created in atomic traps. Interestingly, theorists are exploiting the Maldacena connection between four dimensional gauge theory and ten dimensional string theory to model such a thermodynamic system\cite{ik}. The model is that of a 10-D rotating black hole with Hawking-Beckenstein entropy. It accommodates the low viscosities observed. This should put us on notice that very high energy collisions of nuclei may be more interesting than anticipated from ÒQCD-inspiredÓ logarithmic extrapolations of accelerator data. This is relevant to the analysis of cosmic ray experiments. Enter particle astrophysics.

Conventional astronomy spans 60 octaves in photon frequency, from $10^4$ cm radio-waves to $10^{-14}$\,cm photons of GeV energy. This is an amazing expansion of the power of our eyes that scan the sky over less than a single octave just above 10$^{-5}$\,cm wavelength. Recently detection and data handling techniques of particle physics\cite{md} are reborn in instrumentation to probe the Universe at new wavelengths, smaller than $10^{-14}$\,cm, or photon energies larger than 10 GeV. Besides gamma rays, gravitational waves and neutrinos as well as very high-energy protons that are only weakly deflected by the magnetic field of our galaxy, have become astronomical messengers from the Universe\cite{ro}. As exemplified time and again, the development of novel ways of looking into space invariably results in the discovery of unanticipated phenomena.  For particle physicists the sexiest astrophysics problem is undoubtedly how Nature manages to impart an energy of more than $10^8$\,TeV to a single elementary particle.

Although cosmic rays were discovered almost a century ago, we do not know how and where they are accelerated\cite{sw}. This may be the oldest mystery in astronomy and solving it is challenging as can be seen by the following argument. It is sensible to assume that, in order to accelerate a proton to
energy $E$ in a magnetic field $B$, the size $R$ of the accelerator must encompass the gyroradius of the particle: $R >  R_{\textrm{gyro}} = E / B$, {\it i.e.}\ the accelerating magnetic field must contain the particle orbit. This condition yields a maximum energy
$ E < \Gamma{BR}$ by dimensional analysis and nothing more. The factor $\Gamma$ has been included to allow for the possibility that we may not be at rest in the frame of the cosmic accelerator resulting in the observation of boosted particle energies. Opportunity for particle acceleration to the highest energies is limited to dense regions where exceptional gravitational forces create relativistic particle flows: the dense cores of exploding stars, inflows on supermassive black holes at the centers of active galaxies, annihilating black holes or neutron stars? All speculations involve collapsed objects and we can therefore replace $R$ by the Schwartzschild radius $ R \sim {GM/c}^2$ to obtain $ E < \Gamma {BM}$.

The above speculations are reinforced by the fact that the sources listed happen to also be the sources of the highest energy gamma rays observed. At this point a reality check is in order. Note that the above dimensional analysis applies to the Fermilab accelerator: kGauss fields over several kilometers (covered with a repetition rate of $10^5$\,revolutions per second) yield 1 TeV. The argument holds because, with optimized design and perfect alignment of magnets, the accelerator reaches efficiencies close to the dimensional limit. It is highly questionable that Nature can achieve this feat. Theorists can imagine acceleration in shocks with efficiency of perhaps 1--10\%.

Given the microgauss magnetic field of our galaxy, no structures seem large or massive enough to reach the energies of the highest energy cosmic rays. Dimensional analysis therefore limits their sources to extragalactic objects. A common speculation is that they may be relatively nearby active galactic nuclei powered by a billion solar mass black holes. With kilo-Gauss fields we reach 100 EeV, or $10^{20}$ eV. The jets (blazars) emitted by the central black hole could reach similar energies in accelerating substructures boosted in our direction by a $\Gamma$-factor of 10, possibly higher. The neutron star or black hole remnant of a collapsing supermassive star could support magnetic fields of $10^{12}$ Gauss, possibly larger. Shocks with $\Gamma > 10^2$ emanating from the collapsed black hole could be the origin of gamma ray bursts and, possibly, the source of the highest energy cosmic rays.

The astrophysics problem is so daunting that many believe that cosmic rays are not the beam of cosmic accelerators but the decay products of remnants from the early Universe, for instance topological defects associated with a grand unified GUT phase transition near $10^{24}$ eV. A topological defect will suffer a chain decay into GUT particles X,Y, that subsequently decay to familiar weak bosons, leptons and quark- or gluon jets. Cosmic rays are the fragmentation products of these jets. HERA again revealed to us the composition of these jets that count relatively few protons, {\it i.e.}\ cosmic rays, among their fragmentation products and this is increasingly becoming a problem when one confronts this idea with data.

We conclude that, where the highest energy cosmic rays are concerned, both the accelerator mechanism and the particle physics are enigmatic. There is a realistic hope that the oldest problem in astronomy will be resolved soon by ambitious experimentation: air shower arrays of $10^4 \, \textrm{km}^2$ area (Auger), arrays of air Cerenkov detectors (H.E.S.S. and Veritas, as well as the Magic 17~m mirror telescope) and kilometer-scale neutrino observatories (IceCube and NEMO). Some of these instruments have other missions; all are likely to have a major impact on cosmic ray physics. While no  breakthroughs were reported, preliminary data forecast rapid progress and imminent results in all three areas\cite{ro}.

\begin{figure}[h]
\centering\leavevmode
\epsfig{file=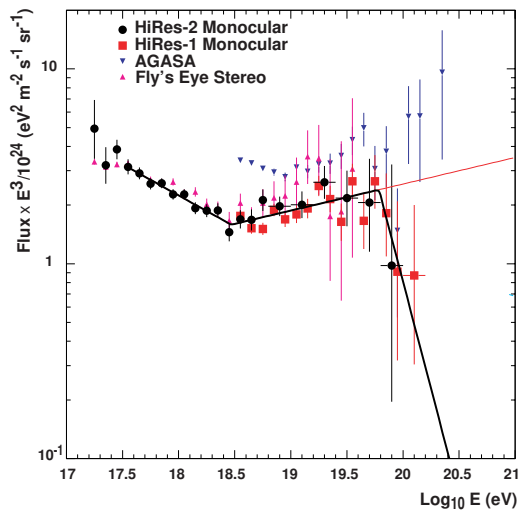, width=2.7in}
\caption{Extragalactic cosmic ray spectrum before Auger and HiRes stereo measurements.}
\label{fig:fluor}
\end{figure}

The Auger air shower array is confronting the low statistics problem at the highest energies by instrumenting a huge collection area covering 3000 square kilometers on an elevated plane in Western Argentina. The instrumentation consists of 1600 water Cherenkov detectors spaced by 1.5 km. For calibration, showers occurring at night, about 10 percent of them, are also viewed by four fluorescence detectors. The detector will observe several thousand events per year above 10 EeV and tens above 100 EeV, with the exact numbers depending on the detailed shape of the observed spectrum. The end of the cosmic ray spectrum is a matter of speculation given the somewhat conflicting results from existing experiments, most notably the HiRes fluorescence detector and the AGASA scintillator array; see Fig.\,\ref{fig:fluor}.

Above a threshold of 50 EeV the cosmic rays interact with cosmic microwave photons and lose energy to pions before reaching our detectors. This is the Greissen-Zatsepin-Kuzmin cutoff that limits the sources to our supercluster of galaxies. The feature in the spectrum is claimed at the 5 sigma level in the latest HiRes data. It is totally absent in the AGASA data, a fact that would require some radical departure from established particle physics or astrophysics. At this meeting Auger presented the first results from the partially deployed array\cite{sw}. The exposure is similar to that of the final AGASA data. The data confirms the existence of super EeV events. There is no evidence, either in the latest HiRes or the Auger data, however, for anisotropies in the arrival directions of the cosmic rays claimed mostly on the basis of the AGASA data. Importantly, Auger observes a discrepancy between the energy measurements of showers obtained from the fluorescence and particle array techniques. The discrepancy suggests that very high energy air showers do not develop as fast as modeled by the particle physics simulations used to analyze previous experiments, {\it i.e.}\ the data necessitate deeper penetration of the primary, less inelasticity and more energy in fewer leading particles than anticipated. Auger data definitely indicate that the experiment is likely to qualitatively improve existing observations of the highest energy cosmic rays in the near future.
 
\begin{figure}[h]
\centering\leavevmode
\epsfig{file=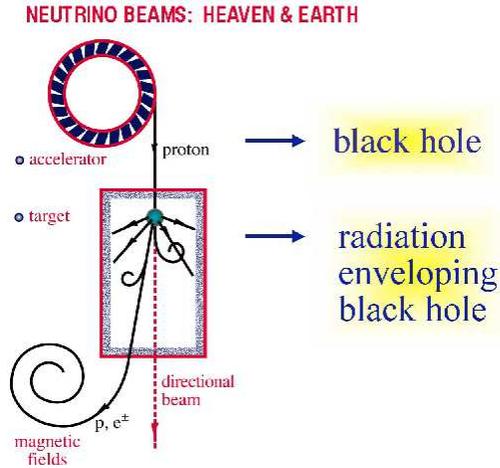, width=2.7in}
\caption{Cosmic ray accelerators are also cosmic beamdumps producing fluxes of neutrinos and TeV photons accompanying the cosmic rays.}
\label{fig:beamdump}
\end{figure}

Cosmic accelerators are also cosmic beam dumps producing secondary photon and neutrino beams\cite{ewax}. Particles accelerated near black holes pass through intense radiation fields or dense clouds of gas leading to production of secondary photons and neutrinos that accompany the primary cosmic-ray beam. The target material, whether a gas or photons, is likely to be sufficiently tenuous so that the primary beam and the photon beam are only partially attenuated; see Fig.\,\ref{fig:beamdump}.

Although gamma ray and neutrino telescopes have multiple interdisciplinary science missions, in the case of neutrinos the real challenge has been to develop a reliable, expandable and affordable detector technology\cite{po}. The South Pole AMANDA neutrino telescope, now in its fifth year of operation, has improved its sensitivity by more than an order of magnitude since reporting its first results in 2000. It has now reached a sensitivity close to the neutrino flux anticipated to accompany the highest energy cosmic rays, dubbed the Waxman-Bahcall bound. Expansion into the IceCube kilometer-scale neutrino observatory, required to be sensitive to the best estimates of potential cosmic neutrino fluxes, is in progress. Companion experiments in the deep Mediterranean are moving from the R\&D into the construction phase with the goal to eventually build an IceCube size detector. With the sun and SN87 neutrino observations as proofs of concepts, next-generation neutrino experiments will also scrutinize their data for new particle physics, from the signatures of dark matter to the evidence for additional dimensions of space.

\begin{figure*}[t]
\centering\leavevmode
\epsfig{file=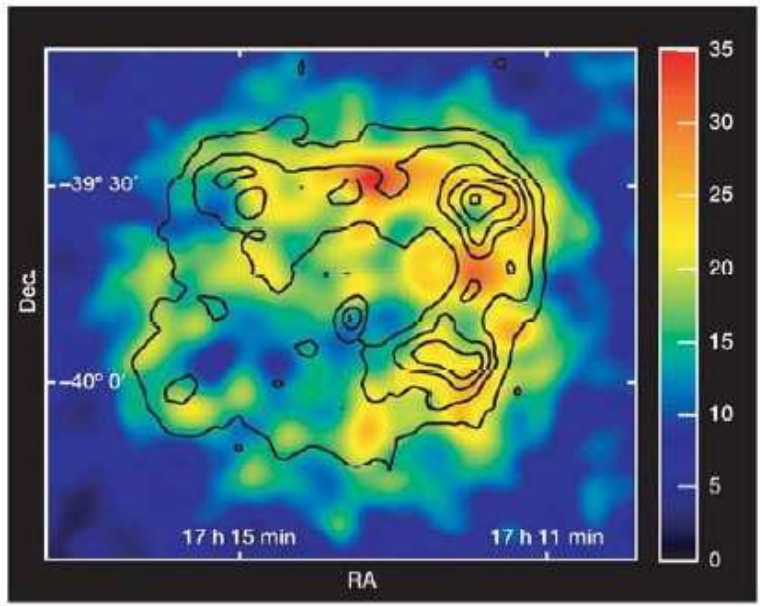, width=4in}
\caption{TeV gamma ray image of a young supernova remnant.}
\label{fig:hess}
\end{figure*}

It is however the H.E.S.S. array of four air Cherenkov gamma ray telescopes deployed under the southern sky of Namibia that delivered the highlights in the particle astrophysics corner\cite{ro}. For the first time an instrument is capable of imaging astronomical sources in TeV gamma rays. Its images of young galactic supernova remnants shows filament structures of high magnetic fields that are capable of accelerating protons to the energies, and with the energy balance, required to explain the galactic cosmic rays; Fig.\,\ref{fig:hess}. Although the smoking gun for cosmic ray acceleration is still missing, the evidence is tantalizingly close.

The big event of the next biennium is the commissioning of the LHC. With dark matter and energy\cite{sh,lb}, astronomers have raised physics problems that seem as daunting as the problem of the lifetime of the sun over one century ago. Evolution and geology required a sun that was older than several tens of millions of years. Chemistry established its lifetime at 3000 years. Neither chemistry nor astronomy solved the puzzle, Rutherford did. May history repeat itself with the solution revealed by the accelerators in our future, LHC and a linear collider\cite{ilc}!

Rendez vous in 2007 in Daegu, Korea.

\section*{Acknowledgments}
I am grateful for the superb hospitality of  Professor Tord Ekelof, my Uppsala IceCube colleagues and other organizers of  the meeting. This talk is not only inspired by the excellent talks at this meeting, but also by a series of stimulating and provocative talks at meetings I attended prior to this conference, by Alan Martin, K. Kinoshita and C. W. Chiang at Pheno 05 in Madison and by Pilar Hernandez, Hitoshi Murayama, Carlos Pena-Garay and G. Raffelt at a neutrino workshop in Madrid. I thank Vernon Barger, Hooman Davoudiasl, Concha Gonzalez-Garcia, Tao Han, and Patrick Huber for reading the manuscript. This research was supported in part by the National Science Foundation under Grant No.~OPP-0236449, in part by the U.S.~Department of Energy under Grant No.~DE-FG02-95ER40896, and in part by the University of Wisconsin Research Committee with funds granted by the Wisconsin Alumni Research Foundation.

\end{document}